# Optical Properties of Ultrashort Semiconducting Single-Walled Carbon Nanotube Capsules Down to Sub-10 nm

Xiaoming Sun,[†] Sasa Zaric,[‡] Dan Daranciang,[‡] Kevin Welsher,[‡] Yuerui Lu,[‡] Xiaolin Li,[‡] and Hongjie Dai*,[‡]

*Department of Chemistry and Laboratory for Advanced Materials, Stanford University, Stanford, California 94305, and Chemistry Department, Beijing University of Chemical Technology, Beijing 100029, P. R. China*

Received January 28, 2008; E-mail: hdai@stanford.edu

**Abstract:** Single-walled carbon nanotubes (SWNTs) are typically long (≳100 nm) and have been well established as novel quasi one-dimensional systems with interesting electrical, mechanical, and optical properties. Here, quasi zero-dimensional SWNTs with finite lengths down to the molecular scale (7.5 nm in average) were obtained by length separation using a density gradient ultracentrifugation method. Different sedimentation rates of nanotubes with different lengths in a density gradient were taken advantage of to sort SWNTs according to length. Optical experiments on the SWNT fractions revealed that the UV−vis−NIR absorption and photoluminescence peaks of the ultrashort SWNTs blue-shift up to ∼30 meV compared to long nanotubes, owing to quantum confinement effects along the length of ultrashort SWNTs. These nanotube capsules essentially correspond to SWNT quantum dots.

## Introduction

Single-walled carbon nanotubes (SWNTs) are rolled up from graphene sheets and have shown unique electronic, optical, and mechanical properties.[1–3] Although the dependence of SWNT physical properties on chirality has been extensively probed both theoretically and experimentally,[1,4] less has been done for length-dependent properties, including optical absorption and photoluminescence. The ultimate miniaturization of SWNTs is to produce nanotube capsules with nanometer-scale lengths. Theory predicts that the bandgap of a nanotube increases with shorter length in an oscillatory manner. The oscillation amplitude increases as the nanotube length decreases. The underlying physics is quantum confinement along the tube length as the nanotube approaches zero-dimensional sizes. This is similar to the quantum size effects observed in other semiconductor systems.[5–9] These changes should be experimentally observable for nanotube lengths below several tens of nanometers.[5–7,10] Scanning tunneling microscope and spectroscopy methods probed metallic nanotubes cut down to 30 nm length and indeed showed the increase in the spacing between energy levels of quantum states confined along the nanotube axis for shorter nanotubes.[11,12]

UV−vis−NIR interband absorption and photoluminescence (PL) spectroscopy methods have proven to be powerful tools for characterizing SWNTs. Results have shown spectral features related to the first and second subband gaps of semiconducting nanotubes present in the sample,[4,13] and for the precise spectral peak positions, exciton effects had to be taken into account.[14] Optical probing of the length regime where strong quantum confinement is expected has not been possible so far because of SWNT sample preparation and separation limitations, although a number of analytical methods have been used for length, diameter, or chirality separation, such as size exclusion chromatography,[15,16] ion exchange chromatography,[17,18] electrophoresis[19] (or dielectrophoresis),[20] isopycnic density gradient separation,[21,22] or combination of these methods.[19,23] The shortest length obtained is typically ∼50 nm. Recently, Fagan et al. obtained nanotube fractions down to 10 nm in length[24] but observed no shifts in SWNT optical spectra as length was reduced.

Here, we report length separation of SWNTs by centrifugal rate separation in a density gradient. We obtain nanotube samples with average lengths down to ∼7.5 nm and reasonable length distribution. The samples were characterized by atomic force microscopy (AFM) and optical absorption and PL-excitation spectroscopy methods. As the length of SWNTs is reduced below 50 nm, the optical spectra show clear blue shifts, demonstrating for the first time that optical peak positions of ultrashort SWNTs are dependent on length because of quantum confinement.

## Results and Discussions

The details of sample preparation and length separation of HiPco SWNTs are given in the Supporting Information (part I). In brief, a layer of 0.2 mL of sonication-cut SWNTs wrapped with surfactant phospholipid (PL) based distearoyl-*sn*-glycero-3-phosphoethanolamine-*N*-[methoxy(polyethylene-glycerol)-5000] (PL-PEG)[25] was put on top of a three-layer water−iodixanol solution density gradient and centrifuged at ultrahigh speed (∼300k$g$) for 3 h (see photograph of the ultracentrifuge tube after separation in Supporting Information, Figure S1A). The gradient containing separated SWNTs was manually sampled and fractioned (100 μL each) from the centrifuge tube for subsequent characterization.

Tapping mode AFM images of typical fractions (see Figure 1) indicated that fraction 6 (labeled as f6 in Figure 1) contained the shortest SWNTs. After AFM tip size correction (see Supporting Information), the SWNTs showed an average length of ∼7.5 nm (longest, 16 nm and shortest, ∼2 nm). The average length of subsequent fractions (f8, f12, and f18) gradually increased to 11, 27, and 58 nm, respectively, with a standard deviation of 45−55%. Length histograms of the various fractions obtained are shown next to the corresponding AFM images in Figure 1.

The method used in our work was different from isopycnic density gradient separation used to separate SWNTs according to diameter and bandgap.[21,22] Multilayer gradient (5, 7.5, 10% layers) was used here, but all layers had a density smaller than that of SWNT for separation; thus, all SWNTs had a chance to go through the gradient, and only the time required to go through the density gradient depended on the length of the nanotubes. By making use of the different sedimentation rates of different length SWNTs in density gradient and by terminating the sedimentation by removing centrifugal force, different length

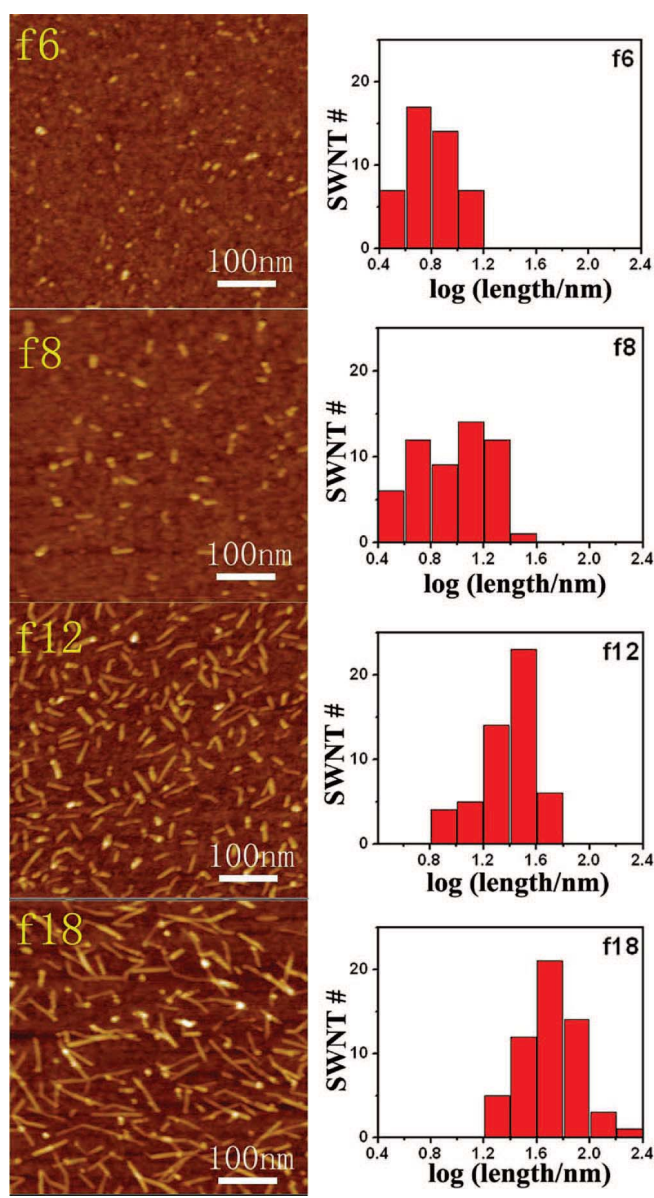

**Figure 1.** AFM images of various SWNT fractions after centrifugal rate separation (f6 means fraction 6 and so on) in gradient. The corresponding length histograms of the fractions measured from 50 tubes in each fraction are shown next to the corresponding AFM images.

SWNTs are captured along the centrifuge tube at different spatial locations. Longer SWNTs are located nearer to the bottom of the centrifuge tube, and shorter SWNTs are nearer to the top. The density gradient needs to be well designed. Isopycnic separation experiment indicated that individual SWNTs wrapped with PL-PEG have a density of ∼1.12 g/cm$^3$, equal to ∼22% iodixanol (Supporting Information, Figure S2). Therefore, the gradient layers used should have a density <22%. As control experiments indicated, iodixanol solution without density gradient could also separate according to length to a certain degree but was poor in performance. Too-sharp an increase in density gradient (e.g., 10%) would make tubes accumulate at boundary areas and impair the separation (Supporting Information, Figure S3A). We found that when the gradient was shallow (e.g., interlayer difference ≤5%) and the centrifugation time was long enough (e.g., ≥2 h), the interlayer boundary became invisible after ultracentrifugation. In that case, the step gradient gave separation results as good as results from the linear gradient. To obtain primarily short tubes, we used a gradient with relatively low density and long centrifugation times. We chose 5 + 7.5 + 10% three-layer step density gradient to effectively control the sedimentation speed of individual SWNTs 2−50 nm in length, thus providing higher resolution in rate separation of short tubes. This design was similar to that used for biomacromolecule separation according to mass.[26] We should note that the rational design on gradient should combine with good timing for centrifugation. Too-long centrifugation time in the above gradient will cause SWNTs with different lengths to settle down near the bottom of the centrifuge tube (Supporting Information, Figure S3B), whereas relatively long SWNTs can be obtained by significantly shortened ultracentrifugation time (Supporting Information, Figure S4A).

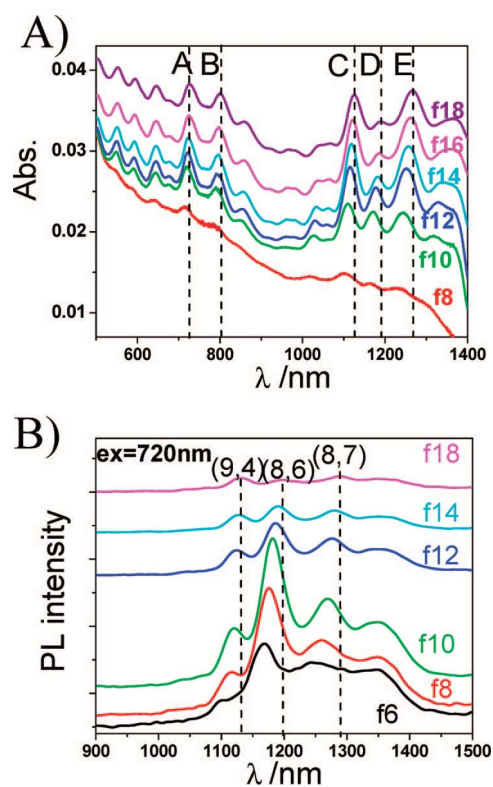

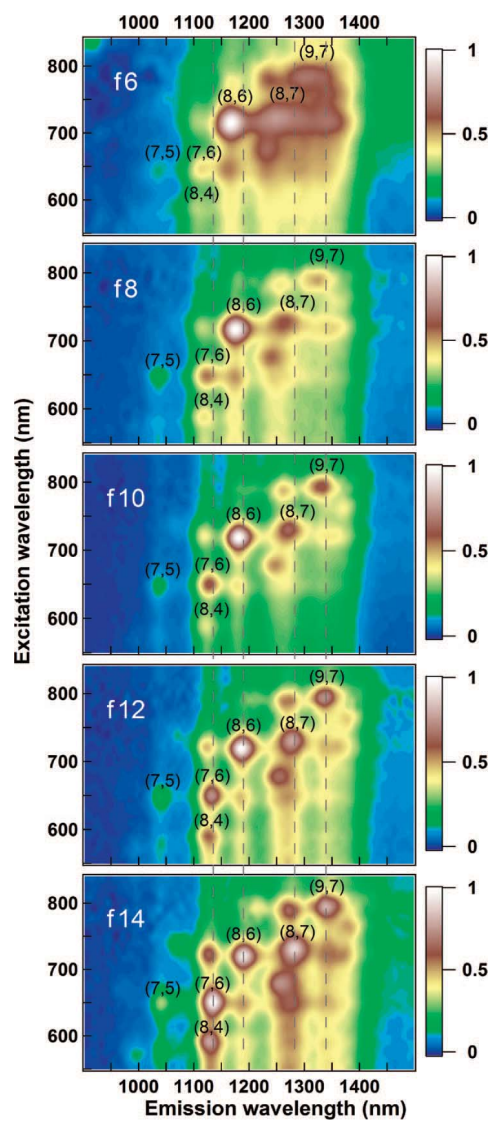

*Figure 2.* Optical characterization of length separated SWNT fractions. (A) UV−vis−NIR absorption spectra after normalization to nanotube concentration (on the basis of absorbance at 935 nm). (B) NIR−PL spectrum under 720 nm excitation after normalization to concentration. Three SWNT chiralities, (9,4), (8,6), and (8,7), were selectively excited. Blue shifts in spectral peaks are seen in both absorption and PL spectra for shorter-length fractions.

The UV−vis−NIR absorption spectra of several fractions are shown in Figure 2A. The absorption peaks in 900−1500 nm range are due to the lowest $E_{11}$ subband absorption, whereas the peaks in 550−900 nm range correspond to their second $E_{22}$ subband transitions[4,13] Multiple peaks corresponding to various semiconducting SWNTs are present in the sample. Both first and second subband absorption peaks show blue shifts as the average nanotube length is reduced. The shift was continual and monotonic: shorter fractions showed stronger blue shift. By using a convenient excitation wavelength in the second subband absorption range, a small number of chiralities are

(26) Price, C. A. *Centrifugation in Density Gradients*; Academic Press: New York, 1982.

*Figure 3.* PLE spectra of various SWNT fractions after length separation. The vertical lines are guide to the eye. Chirality assignments were marked for selected PLE peaks.

selectively excited, and their PL in the first subband range can be measured. Because of the reduced number of overlapping PL peaks compared to the number of absorption peaks of various (*n*,*m*) SWNTs in the first subband region, the measured NIR−PL spectra show better defined spectral features. NIR−PL of various fractions measured by using 720 nm excitation is shown in Figure 2B. The three most prominent peaks correspond to (9,4), (8,6), and (8,7) nanotubes. All three show blue shifts as the fraction average length is reduced, in agreement with the absorption data.

By measuring NIR−PL of various semiconducting (*m*,*n*) SWNTs over a range of excitation wavelengths, length-dependent spectral shifts for a number of semiconducting chiralities present in the sample are clearly gleaned. Figure 3 shows resulting PL excitation (PLE) spectra for several fractions. The PLE peaks were labeled by their chirality assignment.[4] All PLE peaks blue-shifted as the fraction length was reduced both in PL direction and excitation direction (in agreement with shifts observed in first and second subband absorption data, respectively).

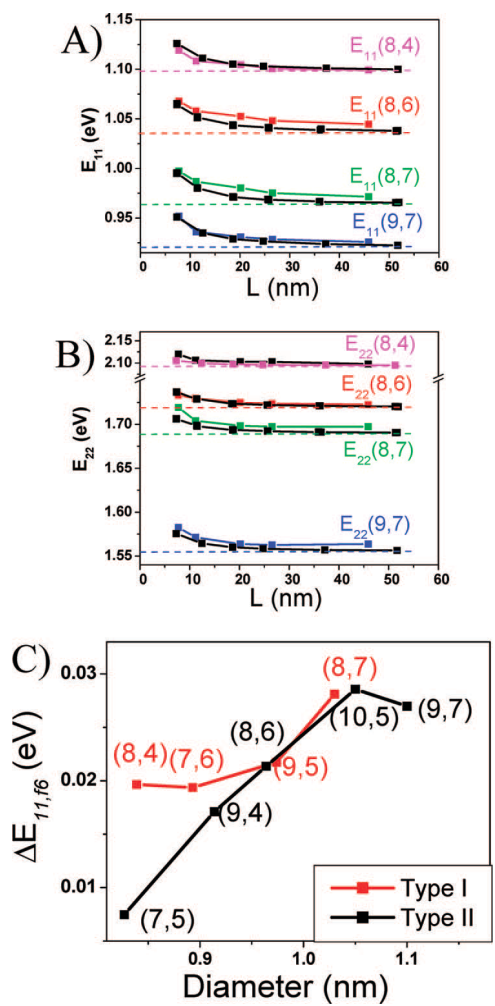

**Figure 4.** PLE peak positions of (8,4), (8,6), (8,7), and (9,7) tubes as functions of tube length. (A) $E_{11}$-PL peak positions vs length. (B) $E_{22}$-excitation peak position vs length. Black curves are modeling results (see Supporting Information, Figure S10). (C) Dependence of the $E_{11,f6}$ shift on tube diameter. For a $(m,n)$ tube, type I means $(m - n)$ mod $3 = 1$, and type II means $(m - n)$ mod $3 = -1$.

Length-dependent PLE peak positions for several chiralities are given in Figure 4A,B (the data for all measured chiralities are given in the Supporting Information, Figure S5). The blue shifts are highly pronounced for nanotube lengths below ∼20 nm. Dotted lines in Figure 4 represent PLE peak positions for long nanotubes (∼140 nm. See Supporting Information, Figure S4), which were obtained through a modified procedure by using shorter sonication for cutting SWNTs and shorter centrifugation time to fractionate long tubes. The peak shifts between PL peak positions for the shortest fraction (f6) and the fraction containing long nanotubes (denoted $\Delta E_{11,f6}$) for all measured chiralities are shown in Figure 4C (excitation peak shifts $\Delta E_{22,f6}$ are given in the Supporting Information, Figure S5). Both $\Delta E_{11,f6}$ and $\Delta E_{22,f6}$ ranged up to ∼30 meV, with the chirality averaged value of 24 and 21 meV, respectively. Note that the $\Delta E_{11,f6}$ and $\Delta E_{22,f6}$ peak shifts were generally smaller for smaller diameter nanotubes.

Before we can assign the observed spectral blue shifts to finite length effects, we need to exclude other reported effects that can lead to spectral peak shifts in SWNTs, such as bundling of nanotubes[13] and the influence of the surrounding medium.[27,28]

The bundling of SWNTs has been shown to cause spectral red shifts[13] and the decrease of PL yield.[29] We measured height histograms of our SWNT fractions by using AFM and identified some bundling in fraction f18 (see Supporting Information, Figure S6), but no bundles were observed in shorter fractions (f6, f8, and f10), where most of the blue shift happened. This was further enforced by calculating relative PL quantum yield from the measured absorption and PLE data, which showed higher PL quantum yield for fractions f6−f10 relative to fraction f18. Because all nanotubes had the same surfactant coating, the only remaining difference in the surrounding medium of different fractions was the percentage of iodixanol used to create density gradient. The difference in iodixanol content between fractions was less than 5%. To verify that iodixanol did not influence the observed peak positions, PLE of HiPco/PL-PEG nanotube samples was measured in the absence and presence of 10% iodixanol, and no shifts were observed (see Supporting Information, Figure S7).

We attribute the observed length-dependent spectra blue shifts of SWNTs to finite length effects. Theoretical modeling (see details of modeling in Supporting Information, Figures S9 and S10) was carried out to obtain a good agreement between the observed spectral shifts and theoretically predicted bandgap changes due to quantum confinement along SWNT length.[6,7,10] The black curves in Figure 4A,B are results of theoretical modeling. Our calculated spectra shifts were based on SWNT bandgap changes for each $(m,n)$ as a function of length and averaged over the length distribution of SWNTs in each fraction (see Supporting Information). A monotonic bandgap change was therefore seen for fractions with decreasing average tube length, without oscillatory behavior between fractions because of averaging over the length distribution. Bandgap change for a $(m,n)$ tube as a function of length was due to increased quantization of states along the nanotube axis. The spectral shifts were more pronounced for larger diameter SWNTs because of the smaller bandgap and higher curvature of the $E(k)$ dispersion curve at the bandgap edge of larger tubes. This was consistent with the observed spectral shifts being more pronounced for larger diameter tubes than for smaller tubes (see Figure 4C). Length distribution in each fraction, together with the increasing bandgap oscillations as the nanotube length is decreased, might have caused broadened PLE peaks, as experimentally observed (see Figure 3 and the experimental length dependence of widths in the Supporting Information, Figure S8). It is also possible that decreased exciton lifetime as nanotube length approaches the exciton size (several nanometers)[14,30,31] caused spectral peak broadening. Interestingly, it is also observed that type-I tubes [i.e., $(m - n)$ mod $3 = 1$] showed spectral shifts larger than those of type-II SWNTs [$(m - n)$ mod $3 = -1$], especially in the small diameter region (≲1 nm, Figure 4C). This was consistent with the fact that type-I SWNTs exhibited smaller $E_{11}$ bandgaps or higher band-edge curvatures than type-II SWNTs of similar diameters because of trigonal warping effects, which are more pronounced for small diameter tubes.[32]

The finite length effects described in this work were not observed previously with separated CoMoCAT nanotubes.[24] We

(27) Moore, V. C.; Strano, M. S.; Haroz, E. H.; Hauge, R. H.; Smalley, R. E.; Schmidt, J.; Talmon, Y. *Nano Lett.* **2003**, *3*, 1379−1382.
(28) Ohno, Y.; Iwasaki, S.; Murakami, Y.; Kishimoto, S.; Maruyama, S.; Mizutani, T. *Phys. Rev. B* **2006**, *73*, 235427.
(29) Crochet, J.; Clemens, M.; Hertel, T. *J. Am. Chem. Soc.* **2007**, *129*, 8058−8059.
(30) Chang, E.; Bussi, G.; Ruini, A.; Molinari, E. *Phys. Rev. Lett.* **2004**, *92*, 196401.
(31) Perebeinos, V.; Tersoff, J.; Avouris, P. *Phys. Rev. Lett.* **2004**, *92*, 257402.

explain this by the smaller shifts observed for smaller diameter nanotubes present in CoMoCAT samples (see Figure 4C) and the shorter nanotube lengths of our samples. Also, the observation of systematic blue shifts requires that other factors, such as bundling and too-wide length distributions, are excluded.

## Conclusion

A new separation method, rate separation in density gradient, was developed to sort 7−60 nm SWNTs acording to length. It led to SWNTs less than 10 nm in length. Optical experiments on the SWNT fractions revealed that the UV−vis−NIR absorption and PL peaks of the ultrashort SWNTs blue-shift up to ∼30 meV compared to long nanotubes, owing to quantum confinement effects along the length of ultrashort SWNTs, essentially corresponding to SWNT quantum dots.

**Acknowledgment** We thank Dr. Jie Deng for helpful discussion. This work was supported by MARCO-MSD and Intel.

**Supporting Information Available:** Experimental details, supplementary figures, and details of modeling of SWNT bandgap vs length. This material is available free of charge via the Internet at http://pubs.acs.org.

JA8006929